\documentclass[10pt,preprint]{aastex}

\usepackage{vmargin}
\usepackage[T1]{fontenc}
\setmargins{1in}{.5in}{6.5in}{9in}{12pt}{30pt}{12pt}{36pt}
\usepackage{epsfig,float}
\usepackage{epsf}
\usepackage{graphicx}
\usepackage{amsmath}
\usepackage{amssymb}
\usepackage{latexsym}
\usepackage{wasysym}
\usepackage{multirow}
\usepackage{url}

\RequirePackage{xspace}
\usepackage{relsize}

\newcommand{\pv}{0.736}
\newcommand{\qv}{0.182}
\newcommand{\meanav}{0.35}

\shorttitle{Characterizing Contaminating Distances}
\shortauthors{Falck, Riess, \& Hlozek}

\begin{document}

\title{Characterizing the contaminating distance distribution for Bayesian supernova cosmology}
\author{Bridget L. Falck\altaffilmark{1} and Adam G. Riess\altaffilmark{2}}
\affil{Department of Physics and Astronomy, Johns Hopkins University, Baltimore, MD 21218}
\and
\author{Ren\'ee Hlozek}
\affil{Department of Astrophysics, Oxford University, Oxford, UK, OX1 3RH}

\date{[Accepted for publication in the Astrophysical Journal]}

\altaffiltext{1}{Electronic address: bfalck@pha.jhu.edu}
\altaffiltext{2}{Space Telescope Science Institute, Baltimore, MD 21218}

\begin{abstract}

Measurements of the equation of state of dark energy from surveys of thousands of Type Ia Supernovae (SNe Ia) will be limited by spectroscopic follow-up and must therefore rely on photometric identification, increasing the chance that the sample is contaminated by Core Collapse Supernovae (CC SNe).  Bayesian methods for supernova cosmology can remove contamination bias while maintaining high statistical precision but are sensitive to the choice of parameterization of the contaminating distance distribution.  We use simulations to investigate the form of the contaminating distribution and its dependence on the absolute magnitudes, light curve shapes, colors, extinction, and redshifts of core collapse supernovae.  We find that the CC luminosity function dominates the distance distribution function, but its shape is increasingly distorted as the redshift increases and more CC SNe fall below the survey magnitude limit.  The shapes and colors of the CC light curves generally shift the distance distribution, and their effect on the CC distances is correlated.  We compare the simulated distances to the first year results of the SDSS-II SN survey and find that the SDSS distance distributions can be reproduced with simulated CC SNe that are $\sim 1$ mag fainter than the standard \citet{rich} luminosity functions, which do not produce a good fit.  To exploit the full power of the Bayesian parameter estimation method, parameterization of the contaminating distribution should be guided by the current knowledge of the CC luminosity functions, coupled with the effects of the survey selection and magnitude-limit, and allow for systematic shifts caused by the parameters of the distance fit.

\end{abstract}

\keywords{cosmological parameters --- dark energy --- supernovae: general}

\section{INTRODUCTION}

Type Ia Supernovae (SNe Ia) can potentially enable the most precise measurement of the equation of state of dark energy at low to intermediate redshifts.  Future ground-based surveys will collect thousands (Pan-STARRS \citep{kaiser}, DES \citep{bernstein}) to millions (LSST \citep{tyson}) of supernova light curves in this redshift range; however, obtaining spectra of such a large number of candidates will be prohibitively expensive.  Thus photometric methods are likely to be used to identify all but only a few percent of transients that are tagged for spectroscopic follow-up.  Many photometric identification methods for supernovae have been proposed~\citep{dahlgoob,poznanski,galyam,johnson,sullivan,kuznetsova,poz07,connolly,rodney,scolnic,gong}, but none purport to be as robust as spectroscopic identification.  Photometrically-identified supernova samples will suffer from both reduced purity (giving how many objects in the sample are actually SNe Ia) and reduced completeness (giving the amount of SNe Ia that survive the various cuts and make it into the final sample).  A sample that is contaminated by Core Collapse (CC) SNe or other objects -- possibly by as little as 2-5\% of SNe Ib/c \citep{homeier} -- will introduce biases in the recovered cosmology parameters, and removing true SNe Ia from the sample will reduce the statistical precision of the measurement.

A high precision measurement of dark energy from a large sample of photometrically-identified supernovae will not be possible unless these issues are addressed. To that effect, a SN photometric classification challenge has been issued in an attempt to test the relative merits of various classification and photo-$z$ estimation methods \citep{kess10}. Additionally, the cosmological measurements of photometrically-identified SN samples can be improved by the Bayesian parameter estimation method developed by \citet{kunz}. This method uses all of the candidate objects, allowing for the possibility that each object belongs to one of two (or more) species, each of which has a different probability distribution describing the apparent distances they are likely to have (at a given redshift) when fit as a SN Ia.  The individual probability that each object is a SN Ia (or another species) is used to discriminate between whether the object should be treated as a Ia, drawn from a narrow distribution, or is a contaminant, drawn from some other distribution.  This is an alternative to imposing cuts on the sample, which will reduce the statistical precision of the measurement and still may not leave the sample 100\% pure, resulting in bias.

The Bayesian method requires some assumption about the probability distribution of the distances of both the Ia and the contaminating CC SNe (which, we should stress, are not real distances).  For SNe Ia, this is the usual likelihood function, where the deviation from the true distance is described by a Gaussian function with a dispersion given by the intrinsic variation of SN Ia luminosities ($\sim 0.15$ mag) and the measurement error added in quadrature (and additionally a dispersion due to the SN redshift uncertainty and peculiar velocity may be added~\citep{kessler}).  The contaminating distribution is not as straight-forward; it will likely include objects of different CC supernova types, each with different luminosity functions (though the method allows for multiple probability distributions), and these ``distances'' are derived by fitting the non-Ia light curves with a set of SN Ia light curve parameters.  This is inherent in the problem, since we do not know ahead of time to which class each object belongs.  

Thus the contaminating distribution at a given redshift will be a complex function of the contaminants' absolute magnitudes, intrinsic colors, extragalactic extinction, light curve shapes, and how these alter the predicted distance when described by a SN Ia model.  Additionally, this distribution may not be the same at every redshift.  Though the contaminating distribution does not need to be known exactly -- it may be parameterized with any number of parameters which are marginalized over at the end -- the choice of parameterization should follow the functional form of the contaminating distances derived for the non-Ia objects.

This paper attempts to characterize the distribution of contaminating distances.  To do this, we run a series of simulations to determine the contaminating distribution as a function of redshift, contaminant supernova type, and the contaminants' absolute magnitudes, extinction, and light curve shapes.  By comparing input parameters to those determined by the distance fitter for a set of realistic simulated light curves, we may determine how each component of the output distance contributes to the final distance function, and how this function changes as the simulated data get noisier at higher redshifts.  The hope is that this study will guide future supernova cosmology endeavors as they attempt to extract the dark energy equation of state out of the thousands to millions of observed supernovae.

\section{METHOD}
\label{sec:method}

In this section we present a brief overview of the Bayesian supernova cosmology framework, motivate our approach to investigating the core collapse distance modulus residual distribution, and describe the simulations used to carry out this study.

\subsection{Bayesian Estimation Applied to Multiple Species}
The presence of non-Ia objects in the sample means that the data is drawn from multiple probability distributions instead of one distribution describing the SN Ia distances (at a given redshift).  \citet{kunz} have developed a method to deal with just this issue, which they call BEAMS: Bayesian Estimation Applied to Multiple Species.  In the BEAMS framework the data are fit using a posterior which weights the likelihood by the probability that each object is a SN Ia
(see~\citet{kunz} for a full derivation):
\begin{equation}
P(\theta|\mu_i) \propto P_i\,\mathcal{L}_{i,\mathrm{Ia}} + (1-P_i)\,\mathcal{L}_{i,\mathrm{non-Ia}},
\end{equation}
with 
\begin{equation}
\mathcal{L}_{i,\mathrm{Ia}} = \frac{1}{\sqrt{2\pi}\sigma_i}e^{-(\mu_i - \mu_\mathrm{th}(\theta))^2/2\sigma_i^2},
\end{equation}
where supernova $i$ has a probability $P_i$ of being type Ia, a measured distance modulus $\mu_i$, an error $\sigma_i$ which includes the measurement error and the intrinsic dispersion of Ia magnitudes added in quadrature, and a redshift $z_i$ which is used to calculate the theoretical distance modulus $\mu_\mathrm{th}(\theta)$ for the cosmological parameters given by $\theta$.  $\mathcal{L}_{i,\mathrm{Ia}}$ is the usual likelihood function for SN Ia distances, weighted by $P_i$, and $\mathcal{L}_{i,\mathrm{non-Ia}}$ is the non-Ia likelihood that describes the contaminating distribution, weighted by $(1-P_i)$.
The parameters describing the contaminating distribution are added to the variables to be fit and marginalized over at the end.  
The probability $P_i$ that candidate $i$ is a Type Ia SN can be calculated by comparing the light curves to templates, or by some other method, and then fed directly into the cosmological parameter estimation instead of being used to make cuts in the sample.  
Alternatively, a global probability parameter may be introduced and marginalized over in the case that the individual probabilities are unknown or uncertain~\citep{kunz,gong}.

\subsection{The Contaminating Distance Function}

The ``distance'' to a contaminant object is determined by fitting a set of multi-band, multi-epoch apparent magnitude light curves to SN Ia templates described by the parameters ($t_\mathrm{max},A_V,\Delta,\mu$), where $t_\mathrm{max}$ is the epoch of maximum light, $A_V$ is the extinction, and $\Delta$ is the MLCS2k2 parameter describing the luminosity-width relation for SNe Ia.
The peak apparent magnitude of a Type Ia supernova template, for observer-frame filter $y$, is given by
\begin{equation}
m_{y,\mathrm{Ia}} = M_{V,\mathrm{Ia}} + \mu_\mathrm{Ia} + P_V\Delta + Q_V\Delta^2 + A_{V,\mathrm{Ia}} + K_{Vy,\mathrm{Ia}},
\end{equation}
while that of an observed CC supernova is given by
\begin{equation}
m_{y,\mathrm{CC}} = M_{V,\mathrm{CC}} + \mu_\mathrm{CC} + A_{V,\mathrm{CC}} + K_{Vy,\mathrm{CC}}.
\end{equation}
The constants $M_{V,\mathrm{Ia}} = -19.5 + 5\log(H_0/65)$, $P_V=\pv$, and $Q_V=\qv$ have been determined from a training set of SN Ia light curves~\citep{jha}; the cross-filter K-corrections~\citep{kgp} are calculated from a set of spectral templates during the fit; and the SN redshift is an input to the fitter.  (Note that throughout this paper we assume we have accurate knowledge of the redshifts, either from spectroscopy of the SN itself or its host galaxy.) 
Rearranging the above two equations and subtracting one from the other, we get the distance modulus residual for a contaminating supernova,
\begin{equation}
\mu_\mathrm{Ia} - \mu_\mathrm{CC} = m_\mathrm{Ia} - m_\mathrm{CC} - (M_\mathrm{Ia} - M_\mathrm{CC}) 
 - (A_\mathrm{Ia} - A_\mathrm{CC})
 - f(\Delta) - (K_\mathrm{Ia} - K_\mathrm{CC})
\end{equation}
(dropping the filter designations), where $f(\Delta) = \pv\Delta + \qv\Delta^2$ and we set $\Delta=0$ for CC SNe.  This distance residual is a function of the true values for each supernova, designated by CC; the best-fit parameters from the distance fitter, $\mu_\mathrm{Ia}$, $A_\mathrm{Ia}$, $\Delta$, (and $t_\mathrm{max}$); and other quantities, such as the K-correction, calculated internally by the distance fitter. Thus the distance residual for a given supernova has several components:
\begin{equation}
\delta\mu = \delta m - \delta M - \delta A - \delta\Delta - \delta K,
\end{equation} 
where $\delta$ denotes the output or calculated value minus the true value. For a set of contaminating CC SNe, the distribution of their distance residuals is exactly the contaminating distribution which must be parameterized in the Bayesian supernova cosmology formalism.

For a well-observed (good signal-to-noise) set of SNe Ia, the distribution of distance residuals is a function of the intrinsic scatter of the Ia luminosities (after accounting for the luminosity-width relation) and the residuals of the other fit parameters, which would be distributed normally about their true values.   
When CC light curves are fit with Ia templates, however, there is little reason to suspect that the residuals of the fit parameters are normally distributed.  For example, the $\delta\Delta$ distribution will likely depend on the distribution of the shapes of the different CC SNe in the sample: $\Delta$ describes the SN Ia width-luminosity relation, whereby brighter ($\Delta < 0$) SNe have broader light curves and fainter ($\Delta > 0$) SNe decline faster~\citep{phillips}.  Additionally, the $\delta A$ distribution will likely depend on the distribution of CC colors, since the color excess is measured with respect to the Ia template color, and it is unlikely that the CC colors are normally distributed about the Ia colors.

We can examine how the distribution of each residual individually affects the distribution of distance residuals by allowing one component to vary while fixing the other residuals to zero.  For $\delta M$ this is done by setting the CC absolute magnitudes to $M_\mathrm{Ia}$ in the simulations, for $\delta A$ this is done by fixing the parameter to the true value in the distance fit, and for $\delta\Delta$ we set $\Delta=0$ (for CC SNe) or the true value (for SNe Ia) in the distance fit.  We do not attempt to fix the $\delta m - \delta K$ component (which we call the zero-point residual), which is caused by the error in the time of peak brightness $t_\mathrm{max}$, the erroneous calculation of the K-corrections, and general random noise in the apparent magnitude light curves.  Correlations may be investigated as we allow multiple components to vary, and we also look at how the residual distributions vary with redshift and CC type.

\subsection{The Simulations}

This section describes the parameters and methods used to produce the simulated observations for this study. During the course of this work the supernova analysis package SNANA\footnote{\url{http://www.sdss.org/supernova/SNANA.html}} \citep{snana} was released which, in addition to simulating supernova surveys, also contains light curve fitting and cosmological analysis software. While SNANA is well-suited to forecasting future surveys and for general analyses of large SN samples, our simulations are tailored to study the specific effects of fitting CC SN light curves to Ia templates as would happen in a contaminated sample of supernovae.

\begin{figure}[htb]
\centering
\includegraphics[angle=90,scale=0.7]{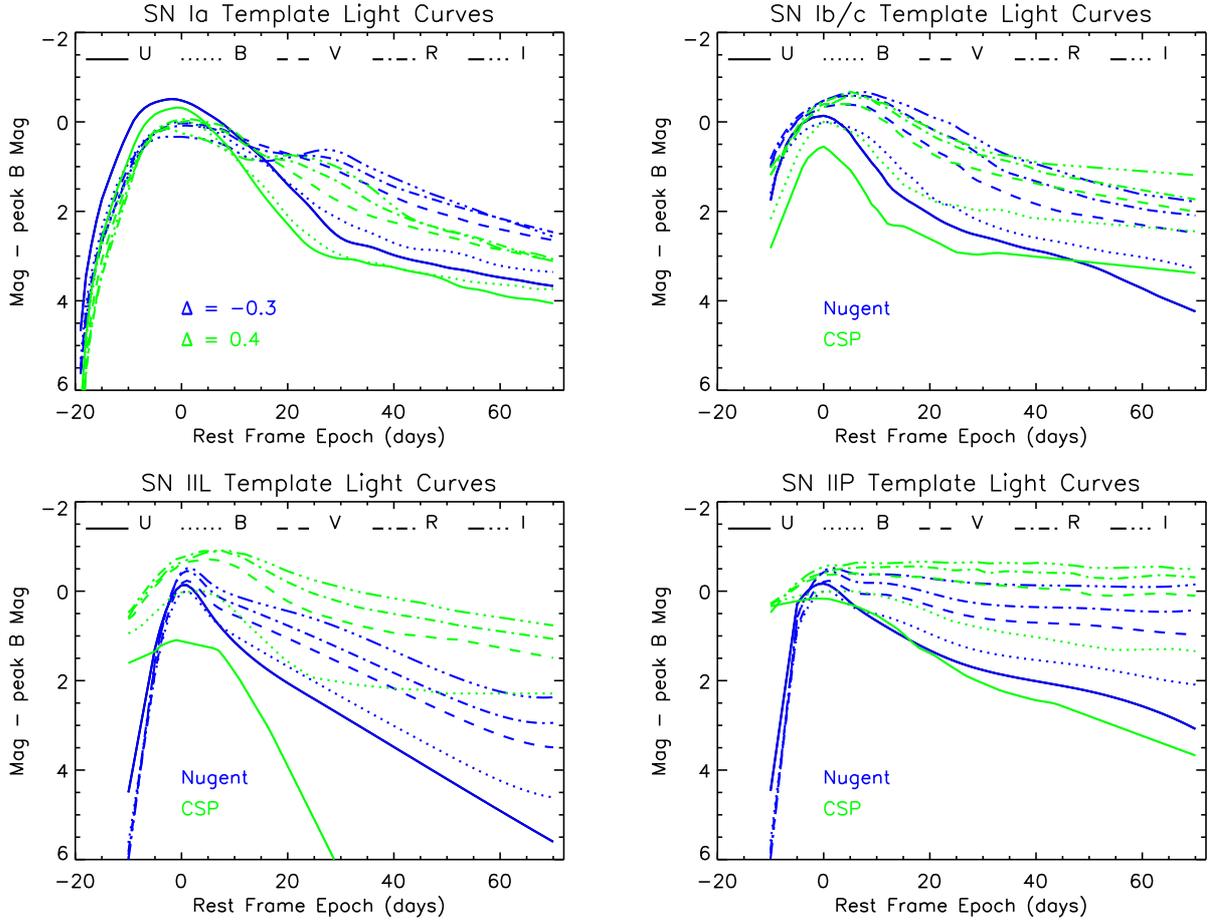}
\caption{Template UBVRI light curves for both Ia and CC SN types.  The SN Ia light curves are shown for $\Delta=-0.3$ (blue) and $\Delta=0.4$ (green), and the CC light curves show both the CSP (green) and Nugent (blue) templates.}
\label{fig:tempLCs}
\end{figure}

We simulate SN observations using the SN Ia Branch-normal, SNIb/c, SN IIL, and SN IIP spectral and light curve templates of P. Nugent\footnote{\url{http://supernova.lbl.gov/nugent/nugent_templates.html}}.
The template SEDs are interpolated between epochs where necessary and are integrated to create rest-frame UBVRI light curves when they do not otherwise exist.  In addition, we create Ib/c, IIL, and IIP templates based on SNe observed by the Carnegie Supernova Project\footnote{\url{http://csp1.lco.cl/~cspuser1/PUB/CSP.html}}.  The Ib/c, IIL, and IIP CSP templates are based on SN2004fe, SN2004ex, and SN2004er respectively; these were chosen because they had well-observed, good signal-to-noise light curves with sufficient pre- and post-peak coverage.  Figure~\ref{fig:tempLCs} shows the rest-frame SN Ia (for two values of $\Delta$), SN Ib/c, SN IIL, and SN IIP (both Nugent and CSP) template light curves.

We give each SN Ia a random luminosity parameter $\Delta$, drawn from an empirical distribution of observed local SNe Ia~\citep{hicken}; to avoid biases in the distribution due to faint (high $\Delta$) objects falling below the magnitude limit of the survey, we define the distribution using only objects with $z \le 0.04$.  Additionally, we add an intrinsic luminosity variation with a dispersion of 0.12 mag and a color variation with a dispersion of 0.05 mag~\citep{nobili}.  These dispersions were chosen such that a simulated sample of well-observed (high signal-to-noise) SN Ia light curves produce distances with a dispersion of about 0.15 mag, matching what is found in the literature.  The CC SN absolute magnitudes, unless they are set to $M_\mathrm{Ia}$, are drawn from Gaussian distributions with means and dispersions taken from~\citet{rich} (see Table~\ref{tab:fracs}); additionally, depending on the specific simulation, the CC SNe may have an added color variation with a dispersion of 0.15 mag~\citep{sullivan}.

\begin{table}[htb]
\centering
\begin{tabular}{cccc}
SN Type & Fraction & $M_B$ (mag) & $\sigma_M$ (mag) \\
\tableline
Ib/c & 0.24 & -17.63 & 1.4 \\
IIL & 0.31 & -17.63 & 0.9 \\
IIP & 0.45 & -16.63 & 1.1 \\
\end{tabular}
\caption{Core Collapse supernova types, intrinsic fractions, and luminosity characteristics.  Fractions are modified from \citet{dahlfran}, which includes sub-populations and Type IIn SNe that we ignore here; $M_B$ is the peak absolute B-band magnitude from~\citet{rich}, rescaled to our $H_0 = 71$ km s$^{-1}$ Mpc$^{-1}$;  and $\sigma_M$ is the dispersion in peak absolute magnitude~\citep{rich}.}
\label{tab:fracs}
\end{table}

Host galaxy extinction is added using to the extinction law of~\citet{ccm} (updated by~\citet{odonnell}). The value of $A_V$ for each SN is drawn from an exponential distribution (unless it is set to a specific value) with an average of~\meanav\ for CC SNe and 0.30 for SNe Ia~\citep{hatano}.  The total-to-selective extinction ratios used are $R_V = 3.1$ for CC SNe and $R_V = 2.3$ for SNe Ia~\citep{smith}. We find that a 35\% change in $R_V$ only shifts the output distances by as much as 0.2\%, thus our conclusions are insensitive to choice of $R_V$.

Finally, we redshift the magnitudes to the observed frame by adding the distance modulus, calculated at each redshift for a fiducial set of cosmological parameters (where $H_0 = 71$ km s$^{-1}$ Mpc$^{-1}$, $\Omega_M = 0.26$, $\Omega_{DE} = 0.74$, and $w = -1$), and the cross-filter K-corrections, calculated as in~\citep{kgp} as a function of redshift and epoch; the rest frame epochs are also dilated by $(1+z)$.  We choose to simulate SNe in the SDSS-II Supernova survey, instead of a larger planned survey such as Pan-STARRS or LSST, so that we may check our simulations against existing data (see Section~\ref{sec:sdss}).  Supernovae are simulated at 8 discrete redshifts: 0.075, 0.125, 0.175, 0.225, 0.275, 0.325, 0.375, and 0.425.  Simulated observations are in the form of SDSS $g$, $r$, and $i$ light curves~\citep{fukugita}, with noise added in flux space to the calculated apparent magnitude light curves.  The signal-to-noise ratio for each observation is calculated using SDSS-II telescope and instrument parameters, and the calculations are calibrated so that they give the stated $5\sigma$ magnitude limit in each filter~\citep{gunn}.  The SDSS-II cadence is also matched, and some fraction of epochs is removed to simulate poor weather, resulting in an average of one observation every 5 days, in each filter~\citep{frieman}.

The simulated supernovae are fit with MLCS2k2 \citep{jha} to determine a distance.  MLCS2k2 fits the multi-color light curves for the parameters $\mu$, $A_V$, $\Delta$, and $t_\mathrm{max}$.  In all cases we fix the value of $R_V$ to match what is done in the SDSS-II SN analysis.  We use an exponential prior on $A_V$, with a mean of 1/3, and a flat $\Delta$ prior, with $-0.4 < \Delta < 1.8$ and tapered $\sigma = 0.1$ Gaussian ends.  We do not include any redshift error, so the redshift we input to MLCS2k2 is the true redshift.  When we want to set the extinction or $\Delta$ residual to zero -- that is, we are looking at the distance residual as a function of the error in other parameters -- we fix its value in the distance fit.  No cuts are made based on the $\chi^2$ of the fit.

\section{RESULTS}
\label{sec:results}

We first look at the effect of each component of the contaminating distance function on the distance residuals in Sections~\ref{sec:zpres} to~\ref{sec:magres}; we then investigate the correlation between $\Delta$ and $A_V$ in Section~\ref{sec:delav}; and finally, we simulate the full distance residual when all the components are non-zero in Section~\ref{sec:fullsim}.  For most cases we simulate 100 SNe per type and redshift; since we also investigate the effect of the survey magnitude limit on the contaminating distance distribution in Section~\ref{sec:fullsim}, we increase this to 500 SNe per type and redshift.
When $\delta M = 0$, we simulate the CC SNe with an absolute V-band magnitude of $-19.5 + 5\log(H_0/65)$~\citep{jha}, where $H_0 = 71$ km s$^{-1}$ Mpc$^{-1}$; when $\delta A = 0$, we fix $A_V$ to its true value in the MLCS2k2 fit; and when $\delta\Delta = 0$, we fix $\Delta$ in the MLCS2k2 fit to be 0 for CC SNe and its true value for SNe Ia.
Unless otherwise noted, we use only Nugent template light curve shapes, and we do not give the CC SNe an intrinsic color variation.  The main exception to this is the case where all residuals are allowed to vary, in Section~\ref{sec:fullsim}.

\subsection{Zero-point Residuals: $\delta\mu = \delta m - \delta K$}
\label{sec:zpres}

\begin{figure}[htb]
\centering
\includegraphics[angle=90,scale=0.7]{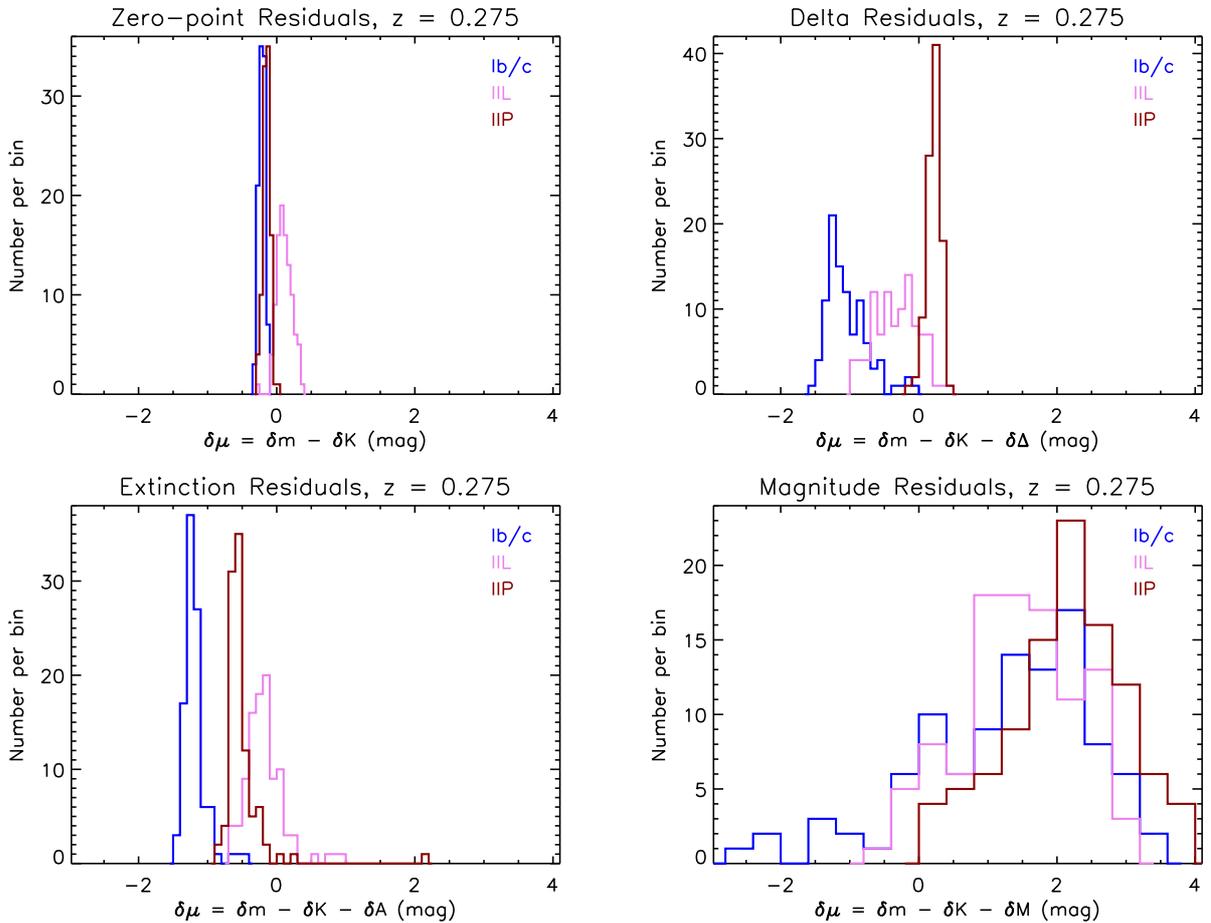}
\caption{The distance residuals caused by the zero-point (top left), $\Delta$ (top right), extinction (bottom left), and absolute magnitude (bottom right) residual distributions at $z = 0.275$.  SNe Ib/c are shown in blue, SNe IIL in violet, and SNe IIP in red.  Note that the y-axes and bin sizes are varied to make each component's histogram clear, but the x-axes are held fixed to stress that the $\delta M$ distribution causes much more variation in $\delta\mu$ than the other components.}
\label{fig:comphist}
\end{figure}

The zero-point residuals are caused by the error in the best-fit time of peak brightness $t_\mathrm{max}$, the erroneous calculation of the K-corrections, and general random noise in the apparent magnitude light curves.  We fix $\delta M = \delta A = \delta\Delta = 0$ as described above, and the distance residual $\delta\mu$ is the best-fitting $\mu$ minus the true distance modulus, which is calculated from the redshift and cosmology.  The distributions of distance residuals are very narrow and centered on a different value for each CC type, and they get wider at higher redshifts from increased noise.  The IIL residual distributions are also wider than the other types at all redshifts, as shown in Figure~\ref{fig:comphist} for $z = 0.275$.
Both the means and widths of the zero-point residual distributions are a strong function of the distribution of output $t_\mathrm{max}$ values, which affect the measured peak apparent magnitudes of the SNe.  The output $t_\mathrm{max}$ of SNe IIP is often several days past the true $t_\mathrm{max}$, but since their light curves are rather flat the difference in apparent magnitude is less; SNe IIL have the widest range of output $t_\mathrm{max}$, giving them zero-point residuals that are fainter than the other CC SNe; and SNe Ib/c have the narrowest light curves, so their output $t_\mathrm{max}$ are well constrained.

\subsection{Delta Residuals: $\delta\mu = \delta m - \delta K - \delta\Delta$}
\label{sec:delres}

In these simulations we allow $\Delta$ to vary -- that is we do not fix $\Delta=0$ in the MLCS2k2 distance fit -- but fix $\delta A = \delta M = 0$.  Since $\Delta$ is a parameter describing the luminosity-width relation of the SN Ia light curves, CC SNe have no such parameter and so the true $\Delta$ is defined to be 0.  The distance residual $\delta\mu$ is then given by the zero-point residual minus the $\Delta$ residual, $\delta\Delta = f(\Delta) = \pv\Delta + \qv\Delta^2$.  As in the previous section, the distributions of the residuals differ for the different CC SNe types, and they get wider as the redshift increases.  They are in general not well-approximated by a Gaussian function (as seen in Figure~\ref{fig:comphist}) and are correlated with the output $\Delta$: lower $\Delta$ means the SN is interpreted as being intrinsically brighter, and the correction to this moves the output distance farther away (fainter) and so increases $\delta\mu$.  It is also interesting to note that the broad widths of Type IIP or ``Plateau'' supernovae cause them to have generally negative output $\Delta$ values.  A large fraction of these have $\Delta < -0.5$, which is outside the range used in the training of MLCS2k2, and such objects are commonly excluded from a sample used to determine cosmology.

\subsection{Extinction Residuals: $\delta\mu = \delta m - \delta K - \delta A$}
\label{sec:avres}

In this simulation we do not fix $A_V$ in the MLCS2k2 distance fit, but fix $\delta\Delta = \delta M = 0$.  The extinction residual, $\delta A$, is the output $A_V$ minus the true $A_V$, which is a random value for each SN drawn from an exponential distribution with an average of~\meanav.  
The output extinction parameters are generally greater than the input $A_V$ ($\delta A > 0$), so $\delta\mu < 0$ and the SNe are interpreted as being nearer than they really are.  This is because in general, CC SNe are redder than SNe Ia at peak~\citep{poznanski}, and so when applying SN Ia templates to fit for CC light curves, this color difference is interpreted as higher extinction.  As in the case of the $\Delta$ residuals, the residual distribution functions are not well-approximated by Gaussian functions, and there is generally evidence for a tail toward the fainter end.

\subsection{Magnitude Residuals: $\delta\mu = \delta m - \delta K - \delta M$}
\label{sec:magres}

In this simulation, $A_V$ and $\Delta$ are fixed in the distance fit, and the absolute magnitude residual $\delta M$ is the SN Ia magnitude, $M_\mathrm{Ia}$, minus the true absolute magnitude.  We give the CC SNe random absolute magnitudes according to the Gaussian luminosity functions in~\citet{rich} (see Table~\ref{tab:fracs}).  Most SNe are thus much fainter than in previous simulations, where the absolute magnitude was set to the characteristic SN Ia absolute magnitude, and most of the SNe at high redshift are too faint to pass the $S/N$ selection criteria (as we will see in Section~\ref{sec:fullsim}).  Figure~\ref{fig:comphist} shows the distance residuals at a median $z$ of $0.275$ for each of the previous components; it is clear that the largest variation in $\delta\mu$ is caused by $\delta M$ due to the wide CC luminosity functions.  It is also notable that each CC type has distinct values for each of the residual components, implying that the contaminating distance distribution will depend on which types of CC SNe are in the final sample used to measure cosmology.

\subsection{Delta and Extinction Correlation: $\delta\mu = \delta m - \delta K - 
\delta\Delta - \delta A$}
\label{sec:delav}

\begin{figure}[htb]
\centering
\includegraphics[angle=90,scale=0.6]{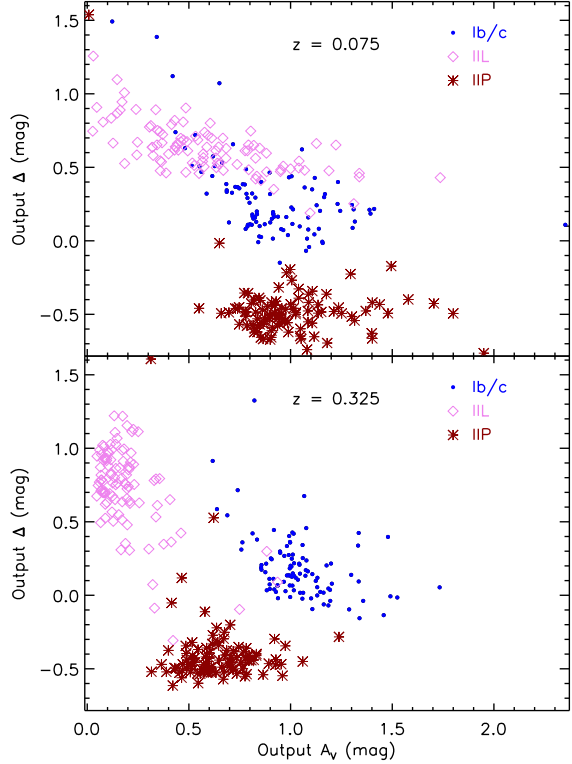}
\caption{The output $\Delta$ vs. output $A_V$ when both are allowed to vary in the distance fit ($\delta\mu = \delta m - \delta K - \delta\Delta - \delta A$), for $z = 0.075$ (top) and $z = 0.325$ (bottom), showing the correlation between $\Delta$ and $A_V$ and its evolution with redshift.  SNe Ib/c are shown in blue, SNe IIL in violet, and SNe IIP in red.  The input $A_V$ values are drawn from an exponential distribution with an average of~\meanav, and the ``true'' $\Delta$ values are 0 for CC SNe.}
\label{fig:delvav}
\end{figure}

When both $A_V$ and $\Delta$ are allowed to vary in the distance fit, the best-fit values of the extinction and $\Delta$ become correlated, and this correlation varies with both redshift and CC type.  Compared to previous $\delta A$, the Ibc and IIP extinction residual distributions are largely unchanged when $\Delta$ is allowed to vary, but the IIL extinction residuals shift toward $\delta A =0$ (e.g. the amount by which MLCS2k2 is overestimating the extinction becomes less).  As seen in Figure~\ref{fig:delvav}, the SNe Ibc and IIL show a trend of decreasing $\Delta$ with increasing $A_V$, while SNe IIP have very low values of the output $\Delta$ and a mean output $A_V$ that decreases with increasing $z$.  SNe IIL show a drastic evolution in the trend with $z$: at low $z$, the range of output $\Delta$ is relatively narrow and positive, and the range of output $A_V$ is wide; at high $z$, the range of output $\Delta$ becomes wide and extends to negative values, while the range of output $A_V$ becomes narrow and low.

The trend in the output $\Delta$ vs. $A_V$ and its evolution with redshift, most apparent for SNe IIL, can be understood to be a result of the dependency of SN Ia max-light colors on the $\Delta$ parameter.  As the best-fit $\Delta$ changes, then, the assumed intrinsic color changes, which will change the value of the measured color excess for a given measured color; if the assumed intrinsic color is bluer, the best-fit extinction increases, and vice-versa.  Figure 8 of \citet{jha} shows the max-light $U-B$, $B-V$, and $V-I$ colors as a function of $\Delta$: as $\Delta$ increases, the $U-B$ and $B-V$ colors increase rapidly while $V-I$ is much more flat.  Thus if the light curve is fit to have a high value of $\Delta$, the redder $U-B$ and $B-V$ SN Ia colors mean the color difference between CC and Ia SNe is less and the best-fit extinction is less, which matches the observed trend in Figure~\ref{fig:delvav}. 

This relationship between $\Delta$ and $A_V$ remains when the CC SNe are simulated with an intrinsic color variation, where the magnitudes in one filter are given a random shift with respect to the magnitudes in another.    
When such a dispersion is added to the CC light curves, there is about 0.1 mag of dispersion added to the distribution of distance residuals.

\subsection{All Residuals: $\delta\mu = \delta m - \delta K - \delta\Delta - \delta A - \delta M$}
\label{sec:fullsim}

We now allow all components of the distance residuals to vary.  In addition, in this simulation we give each SN a random intrinsic color variation, such that the dispersion in color is 0.15 mag~\citep{sullivan}, and we also add variation in the light curve shapes by introducing the CSP templates in addition to the Nugent templates, such that each individual supernova has an equal chance of using either template.  This is an attempt to characterize a wider range of output values than is possible with one set of templates, since a real supernova survey will observe the full gamut of SN light curve shapes, colors, etc.

The results of the previous sections are largely unchanged: each SN type has distinct values of the various residual components, the output $\Delta$ and $A_V$ show the same correlations, and the $\delta M$ distribution continues to dominate the distribution of $\delta\mu$ while the other components cause $\delta\mu$ to shift with respect to $\delta M$.  Additionally, some of the components show a bi-modality, where the CSP SNe are separated from the Nugent SNe by as much as a few tenths of a magnitude.  This suggests that the SN light curve shapes and colors have a strong impact on the best-fitting parameters in the distance fit and is a warning that the Nugent templates alone (plus Gaussian random colors) do not represent the full range of observable SNe; however, the small differences in some of the component residuals are washed out in the $\delta\mu$ distribution due to the large variation in absolute magnitudes.

To mimic the observational effects of a real supernova survey, we require that the supernovae have at least one epoch in each filter with signal-to-noise $S/N  > 5.0$.  This removes simulated supernovae that are not likely to be observed because they are in the noise, and additionally applies a minimal quality cut to the survey.  Such quality cuts attempt to maintain a high precision in the final cosmological analysis.

\begin{figure}[htb]
\centering
\includegraphics[angle=90,scale=0.7]{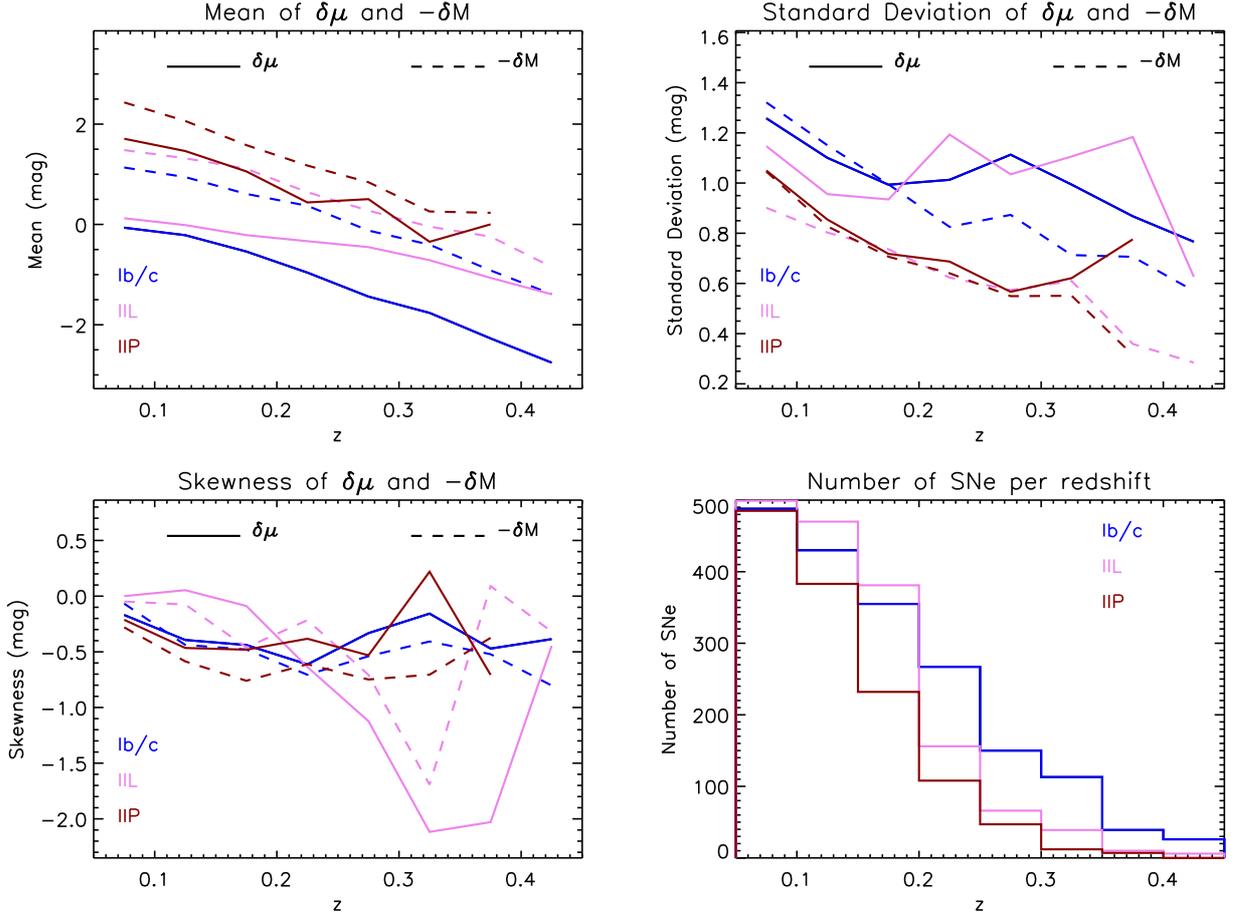}
\caption{The mean, standard deviation, and skewness of the distance ($\delta\mu$; solid) and absolute magnitude ($-\delta M$; dashed) residual distributions as a function of redshift for all 3 CC types, as well as the number of SNe per redshift that pass the signal-to-noise cut.  SNe Ib/c are shown in blue, SNe IIL in violet, and SNe IIP in red.  Note that no SNe IIP pass the cut at the highest redshift, so there are no SNe IIP statistics at this redshift.  The distance residual distribution is a strong function of the magnitude distribution for SNe Ib/c and IIP, but they seem to be less correlated for SNe IIL.}
\label{fig:statsnr}
\end{figure}

When $S/N$ cuts are applied to SNe Ib/c, IIL, and IIP, the $\delta\mu$ distributions of all three SN types show a drastic evolution with redshift.  The output $A_V$ of the SNe that survive at high redshift continue to be rather high while the input $A_V$ decreases.  The mean absolute magnitude gets brighter and the mean distance decreases (gets closer) with redshift, for all SN types, as more and more intrinsically faint SNe fail to pass the $S/N$ cut.  This selection effect also causes the standard deviation of the distance and magnitude residuals to decrease with redshift.  Figure~\ref{fig:statsnr} shows the mean, standard deviation, and skewness of the distance and magnitude residual distributions as a function of redshift, for all 3 CC types.  We discuss each type in turn below.

\begin{itemize}
\item The mean of the SN Ib/c {\it (shown in blue)} distance and magnitude residuals decrease with about the same slope (the other components produce an offset between them), and their standard deviations are also similar, with the distance distribution becoming slightly more spread relative to the magnitude distribution at higher redshifts.  The skewness of the distance residual distribution is slightly negative, and its evolution with redshift also follows that of the absolute magnitude distribution.  This suggests that for SNe Ib/c, the distribution of contaminating distances can be parameterized by the absolute magnitude luminosity function (for the case when this function is Gaussian with a large width) plus the effects of the survey magnitude limit.  

\item The mean of the SN IIL {\it (shown in violet)} absolute magnitude residuals decreases a bit more steeply than that of the distance residuals, due to the changing extinction residuals.  Note however that in the highest redshift bins only a few out of the initial 500 SNe IIL have passed the $S/N$ cut, as they are less likely to have very bright objects than are SNe Ib/c given the simulated luminosity functions.  Because the width of the SN IIL extinction residual distributions are larger than that of the SN Ib/c, and the width of Richardson's SN IIL luminosity function is smaller than that of the SN Ib/c luminosity function, the distribution of contaminating distances and the absolute magnitude luminosity function (after accounting for selection effects) are not as similar for SNe IIL as they are for SNe Ib/c.

\item The means of the SN IIP {\it (shown in red)} distance and absolute magnitude residuals decrease with about the same slope and their standard deviation and skewness are very similar until $z\sim 0.275$, at which point the number of surviving SNe is small and statistics become very uncertain.  Thus as long as SNe IIP make it into the sample, it appears that the distribution of contaminating distances may be parameterized by the absolute magnitude luminosity function, if it is as wide as in~\citet{rich}.
\end{itemize}

It is also general practice to remove SNe from the sample whose light curves have fitted values that lie outside the range used to train the distance fitter.  Thus we also look at the effects of removing all SNe with an output $\Delta < -0.4$ and at the same time applying the above $S/N$ cut.  This mainly affects the SNe IIP, as the SNe Ib/c and SNe IIL have only a few SNe with an output $\Delta$ below this cutoff at all redshifts.  As mentioned previously, SNe IIP are very broad, which gives them very negative values of output $\Delta$.  For the lowest redshift, $z=0.075$, about 80\% of the SNe IIP are removed; this increases to 96\% by $z=0.225$.  Below this redshift, the previous result remains the same, with the mean, standard deviation, and skewness of the distance residuals following that of the absolute magnitude residual distributions.  This suggests, however, that due to their very broad light curves, combined with their faint absolute magnitudes, SNe IIP may not be a significant contributor to the contamination of the SN Ia cosmology analysis.  Thus the contaminating distance distribution will depend mostly on the SN Ib/c and SN IIL luminosity functions and the relative rates of theses CC SN types, combined with the effects of the survey magnitude-limit and candidate selection.


\section{COMPARISON TO SDSS DATA}
\label{sec:sdss}

We now address the question of whether the features of the observed SDSS SNe match those of the simulated SNe and can be explained by the main results presented above.

In addition to the CC SNe presented in Section~\ref{sec:fullsim}, we simulate SNe Ia as described in Section~\ref{sec:method}, varying the $\Delta$, $A_V$, absolute magnitude, and colors and applying the same selection cuts.
To determine how many of each type of SN we would expect to observe, we calculate the number of SN explosions per day per $\deg^2$ as a function of redshift from the intrinsic SN rates per year per comoving volume element: for CC SNe, we use the rates from~\citet{dahlen} and split them into the CC subtypes using the fractions given in Table~\ref{tab:fracs}; and for SNe Ia, we use the rates from~\citet{dahlen08}.  We then multiply this by the fraction, as a function of redshift and type, of SNe that we expect to pass the $S/N$ requirements based on the results of our simulations (i.e. see the bottom right panel of Figure~\ref{fig:statsnr}; note that the CC component is dominated by type Ib/c above $z = 0.25$).  The number of supernovae per redshift for all SN types is normalized in such a way as to produce the best match between data and simulations after all cuts are applied.

To compare the simulations to the SDSS data, we use the first-year sample of identified SNe Ia and CC SNe, as well as unidentified (spectroscopically) SNe, with the requirement that the SN has a spectroscopic redshift from the host galaxy or the supernova itself to avoid any biases caused by photometric redshift errors~\citep{hlozek,smith}.  These SNe satisfy the following conditions: observations on at least 5 different epochs; at least one epoch with $S/N > 5$ in each of the $g$, $r$, and $i$ bands; at least one observation at least 2 days {\it before} peak light, in the rest-frame; and at least one observation at least 10 days {\it after} peak light~\citep{dilday}.  (Note that the simulated SNe are required to satisfy the same conditions.)  Adding the requirement that $\Delta > -0.4$ removes 40-50\% of the SDSS SNe in the lowest two redshift bins; since it is our goal to compare the core-collapse distributions of the SDSS data to the simulations, we do not apply this cut to either the SDSS data or the simulations.  

After adjusting the SDSS distances to have the same value of $H_0$ that is used in the simulations, we calculate the distance residual of each SN by subtracting the distance modulus at the SN redshift (calculated using the same fiducial cosmological parameters as the simulations) from this adjusted distance.  Since the true cosmological parameters may not match those used in the simulations, there may be some small shift in residual space between the SDSS and simulated distance residuals.  (For example, a 15\% change in $\Omega_M$ or a 10\% change in $w_0$ would shift the SDSS distance residuals by $\sim 0.005$ mag at $z=0.05$ and by $\sim 0.03$ mag at $z=0.45$, and a 10\% change in $H_0$ would shift them by $\sim 0.22$ mag at all redshifts.)  We combine the SDSS SNe into 8 equal-width redshift bins, using the discrete redshifts of the simulated SNe as the bin centers.  

\begin{figure}[htb]
\centering
\includegraphics[angle=90,scale=0.7]{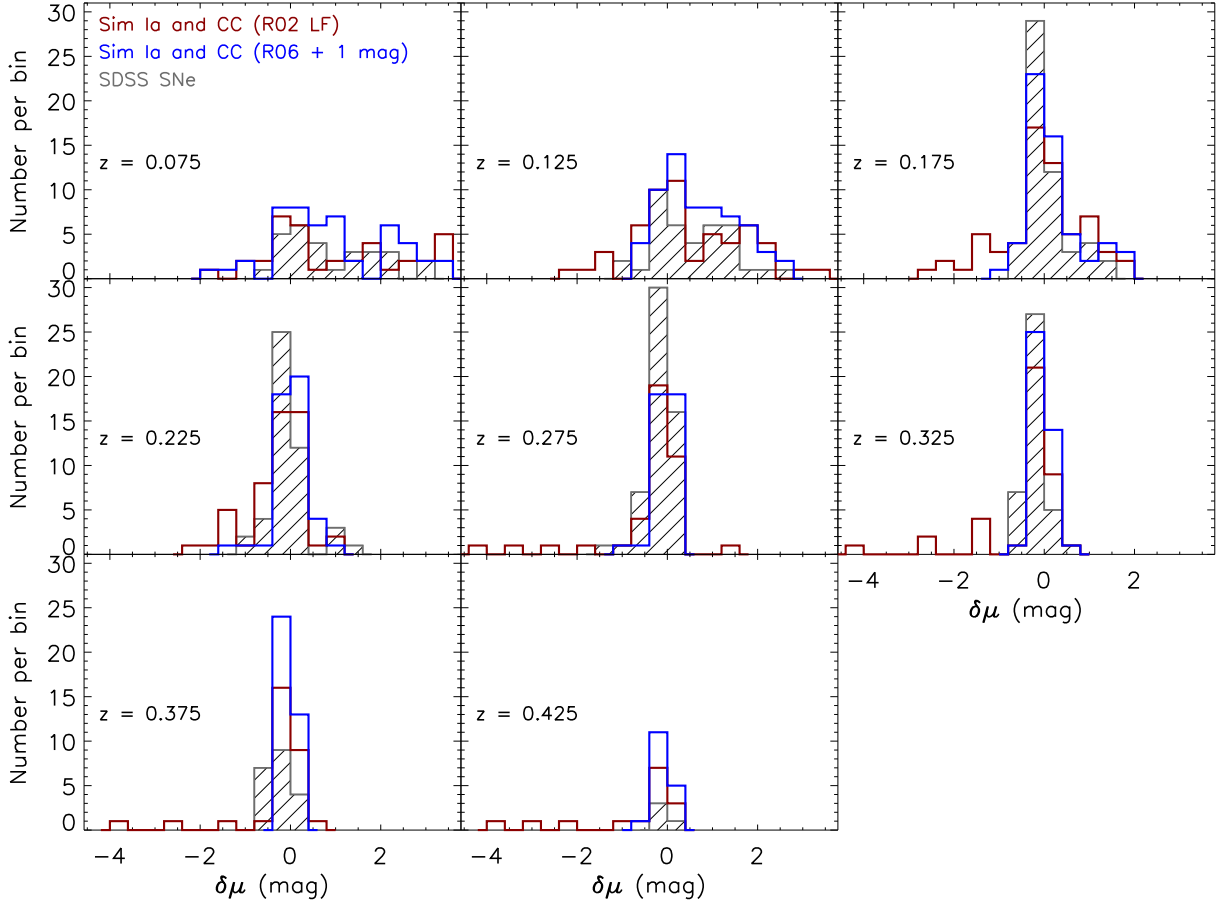}
\caption{The $\delta\mu$ distributions for the 8 $z$ bins.  The SDSS histograms are hashed; the original simulation, using the absolute magnitudes of \citet{rich}, are in red; and the simulation using the narrower Ib/c luminosity function of \citet{rich06}, as well fainter CC absolute magnitudes, are in blue.  There is good agreement between the SDSS residuals and the fainter simulation, while the brighter simulation contains large negative outliers at $z \ge 0.275$.\label{fig:sdvsimagr}}
\end{figure}

The simulated and SDSS distance residuals are in general agreement, but the simulated SNe contain outliers of very close CC SNe, with $\delta\mu < 0$, that do not appear in the SDSS data.
After investigating several possibilities, discussed below, we determine that it is most likely that the absolute magnitude distributions used to simulate the CC SNe are too bright by $\sim 1$ mag, and we re-run the simulation with fainter CC SNe to find better agreement (see Figure~\ref{fig:sdvsimagr}).

{\bf (a) CC vs. Ia rate too high:}
The excess of bright CC SNe causing large negative distance residuals may be a result of including too many simulated CC SNe with respect to SNe Ia; this would mean that while these bright SNe occur, they are rarer than assumed for the simulated data and so are not showing up in the SDSS data.  The ratio of the core collapse to Ia rate used to populate the simulation distance residuals is $R_\mathrm{CC}/R_\mathrm{Ia} = 3.6$ at $z = 0.3$. \citet{botti} determines $R_\mathrm{CC} = 1.15$ at $z = 0.21$ and $R_\mathrm{Ia} = 0.34$ at $z = 0.3$; using their Figure 10 to extrapolate the CC rate to $z = 0.3$, we estimate that $R_\mathrm{CC}/R_\mathrm{Ia} \sim 4$ at $z = 0.3$.  \citet{bazin} finds an even larger value of $R_\mathrm{CC}/R_\mathrm{Ia} = 4.5$ at $z \sim 0.3$.  Thus the relative CC to Ia rate used in the simulations is less than other values in the literature and cannot cause the excess of bright simulated CC SNe.

{\bf (b) Fraction of Ib/c too large:}
If the relative number of simulated SNe Ib/c with respect to SNe II is too high, that would contribute to there being an excess of bright CC SNe that is not observed in the SDSS data, since the SNe Ib/c are more likely to be very bright than are SNe IIL or IIP.  
The fraction of SNe Ib/c with respect to SNe II we use is 24\% (see Table~\ref{tab:fracs}), while other studies give 29\%~\citep{smartt}, 26.5\%~\citep{li}, 24.7\%~\citep{vandenbergh}, 24.6\%~\citep{prieto}, 22.3\%~\citep{cap99}, and 24\%~\citep{bazin}.  Though there is quite a bit of uncertainty in the relative fractions of the CC subtypes, the Ib/c fraction we use is consistent with the literature.

{\bf (c) CC SNe too red:}
If the simulated CC SNe are redder than the SDSS CC SNe, this would contribute to the discrepancy between the simulated and SDSS distance residuals, since this extra reddening would be interpreted as an effect of dust thereby increasing $\delta A$ and decreasing $\delta\mu$.  To investigate this possibility we look at the difference between the rest-frame max-light colors of our simulated CC SNe and those of the template SNe Ia used by the distance fitter, and we compare these color differences to those in~\citet{poznanski} (acknowledging that the available data on CC colors is limited).   
We conclude that the Nugent CC and CSP Ib/c templates provide reasonable representations of CC colors, while the CSP IIL template is quite red, especially in the bluer bands, and the CSP IIP template is only slightly redder than the colors given in~\citet{poznanski}.  Since the simulation outliers are composed equally of CSP and Nugent templates and dominated by SNe Ib/c, they cannot be caused by overly-red CC SNe.

{\bf (d) SDSS SN candidate selection:}
We must also remember that the SDSS-II SN survey follow-up strategy was designed to target SNe Ia preferentially~\citep{sako,kessler}, so it is possible that the SDSS candidate selection process excludes (bright) CC SNe which remain in the final simulated sample.  In order for a SN candidate to make it into the SDSS sample we are considering, it must have a spectroscopic redshift from either the host galaxy or the SN itself. SN Ia candidates (SNe that are likely to be Ia) are identified by matching the available photometry to templates, and these are given the highest priority for spectroscopic follow-up~\citep{sako}. Host-galaxy spectroscopic redshifts were either obtained from the SDSS-I galaxy redshift survey or further follow-up observations of the host galaxies of $\sim$~80 high quality Ia-like candidates~\citep{sako,dilday}, which again selects against CC SNe having spectroscopic redshifts.  

Though we can't simulate exactly the spectroscopic follow-up decisions, we should point out, however, that almost all bright transients with $r < 20.5$ mag {\it were} targeted for spectroscopic follow-up observations~\citep{sako}, so any CC SNe that are this bright would have appeared in the SDSS data.  We find 13 SNe Ib/c, 11 SNe IIL, and 4 SNe IIP that meet this criteria in the simulated sample (and 10, 10, and 3 respectively after applying a $2\sigma$ truncation to the absolute magnitudes), and these have the largest negative distance residuals -- this is a clue that the simulated CC magnitudes are too bright.  

{\bf (e) CC absolute magnitudes too bright:}
The CC luminosity functions we used are from~\citet{rich} (hereafter R02; see Table~\ref{tab:fracs}); due to the limited number of SNe used in that paper to build the absolute magnitude distributions, there is considerable uncertainty in both the mean and width of the CC LFs, and indeed whether they can/should be fit with Gaussian functions.  The simulated distances would come out closer than the SDSS data both if the mean of the distributions were too bright and if the width were too large.  As to the mean, R02 acknowledge that there may be a luminosity bias in their CC distributions, since fainter SNe, if they exist, are less likely to be observed than the brighter SNe.  Their SNe Ia luminosity function is most likely complete, both because SNe Ia are brighter and because R02 find many faint outliers.  As to the width of the absolute magnitude distributions, \citet{rich06} (hereafter R06) look at the absolute magnitudes of stripped-envelope, or Type Ib/c, SNe and find the same mean absolute magnitude but a much reduced width of 0.9 mag, as compared to 1.4 mag in R02.  Again, as in R02, R06 note that there may be a luminosity bias in their sample for which they do not attempt to correct.

We test whether this discrepancy is the result of unexpected absolute magnitudes in the randomly-generated distribution by removing simulated SNe with absolute magnitudes greater than $2\sigma$ from the mean, where $\sigma$ is the R02 width for each CC type.  We find that the furthest outliers are removed from the simulated sample, but outliers remain that are 1-2 mag less than the SDSS residuals.

To estimate the amount by which the simulated absolute magnitudes may be too bright, we look at a recent sample of CC and Ia SNe from SNLS~\citep{bazin}.  In particular, they define a ``pseudo-absolute magnitude'' $\Delta M_{570}$, which is proportional to the absolute magnitude according to the supernova's distance but ignoring the effect of dust absorption.  Figure 11 of \cite{bazin} shows the distribution of $\Delta M_{570}$ both before and after correcting for the detection efficiency and volume of the SNLS survey; after this correction, the CC component seems to be 2-2.5 mag fainter than the Ia component, and there is an increase in the number of events with $\Delta M_{570} > 4$ mag fainter than the SNe Ia that may account for the fainter SNe IIP.  In R02, however, the difference between the mean absolute magnitudes of SNe Ib/c/IIL and SNe Ia is only 1.4 mag.  From this we conclude that the simulated CC SNe could be $\sim 1$ mag too bright; however, since there remain simulated CC distance residuals which are greater than 1 mag removed from the SDSS sample, it is also possible that the Ib/c width of 1.4 mag should be reduced to the R06 width of 0.9 mag.  

To determine if a sufficient match can be made between SDSS data and simulations with fainter CC SNe, we repeat the simulation after adding 1 mag to the mean of the R02 absolute magnitudes and decreasing the width of the Ib/c luminosity function to have the R06 value.  As expected, the outliers caused by very bright CC SNe no longer appear in the simulated distance distributions; indeed, the shapes of the simulated and SDSS residual distributions are very similar (see Figure~\ref{fig:sdvsimagr}).  No CC SNe of any type pass the $S/N$ cut for $z \ge 0.375$, which is consistent with the SDSS data: of the 292 SNe in the SDSS first-year BEAMS sample, only 57 are not categorized as either confirmed Ia (from spectra) or likely Ia (from photometry or low $S/N$ spectra)~\citep{sako,smith}, and all of these 57 have $z < 0.35$.  Though this does not mean that we can say whether the fainter luminosity functions are the true CC luminosity functions, the agreement between data and simulations is encouraging, and it suggests that we may be able to use the simulations to parameterize the contaminating distance distribution for Bayesian supernova cosmology.

\section{CONCLUSIONS}
 
Through the use of detailed simulations of the photometric observations of CC SNe, we have attempted to characterize and account for the features of the distribution of their distances as determined by the Type Ia MLCS2k2 distance fitter\footnote{Note that the main results would be the same but the specifics of how the CC distance function depends on the various component parts (apart from $\delta M$) would change if a distance-fitter other than MLCS2k2 is used.} in order to understand how CC SNe contaminate a sample of SNe Ia distances. 
We have found that the contaminating distance distribution for a given CC type can be characterized by its luminosity function plus a shift in magnitude due to the other components of the distance fit, and its evolution with redshift is determined by the selection effects of a magnitude-limited survey.  The full distribution of contaminating distances will depend on the relative amounts of the CC types, and its evolution with redshift will be affected by the redshift-dependence of the intrinsic CC rates.  Finally, the details of the shape of the contaminating distance distribution will depend on the various selection cuts and qualifiers that go into building the SN sample used to calculate the cosmological parameters.  Thus to better inform the contaminating likelihood function we need to know:
\begin{enumerate}
\item the SN Ib/c, IIL, and IIP absolute magnitude luminosity functions;
\item the relative rates of the CC types; and
\item the selection efficiencies as a function of redshift, SN type, etc. of the survey.
\end{enumerate}
It is not necessary that all of these quantities be known exactly, since the contaminating distribution may be parameterized in the Bayesian framework and the parameters marginalized over, but the choice of parameterization will act as a strong prior on the final results.  It is also possible to solve for the parameters describing the contaminating distribution instead of the cosmology, given a large sample of SN distances, though again the results will depend strongly on the parameterization scheme.

One avenue of future work is to investigate whether the contaminating distribution can be characterized in a non-parametric way by the simulated $\delta\mu$ distributions themselves.  Parameters such as a global shift in magnitude may be added and marginalized over, but the simulations would provide the width and shape of the contaminating distribution. This would be similar to modeling the survey selection efficiencies in spectroscopic SN Ia samples (see, for example, \citet{kessler}). We would need to have confidence that the simulations accurately represent the probability distribution that the data will be drawn from (as a function of distance and redshift), or that the distribution may be varied by a free parameter which is marginalized over without this introducing any biases.

Finally, we would like to stress that these results depend on having accurate redshifts.  Photometric surveys that rely on photometric redshifts instead of host-galaxy spectroscopic redshifts will need to worry about how the redshift errors will affect the contaminating distance distribution, as well as the distances of the SNe Ia and the derived cosmological parameters.  One recent paper~\citep{gong} addresses this question by fitting simulated Ia and CC SNe to a set of Ia templates to determine simultaneously the redshift and the distance, and they use the BEAMS method to determine the recovered cosmology.  We note, however, that they do not take into account the full CC luminosity functions, which we have found to be the dominant component of the contaminating distance distribution, nor do they simulate the selection effects of a magnitude-limited survey, which we have found to affect the evolution of the CC distance distribution with redshift.  Thus a more complicated contaminating distribution function will be needed.

We hope that by highlighting the dominant features affecting the contaminating distribution function, this work will improve both the analyses of future supernova cosmology surveys and the simulations of their effectiveness.  We recommend that future surveys strive to obtain representative spectroscopic follow-up, targeting the full gamut of observed SNe, in order to better understand the contaminating population that will affect any photometric supernova cosmology measurement.

\acknowledgments

We would like to thank Mat Smith for MLCS2k2 fits of the SDSS-II SN data and the anonymous referee for helpful comments.  RH would like to thank Bruce Bassett and Martin Kunz for useful and insightful discussions and the Johns Hopkins University for kind hospitality while working on this project.  RH acknowledges funding from the Rhodes Trust.

Funding for the SDSS and SDSS-II has been provided by the Alfred P. Sloan Foundation, the Participating Institutions, the National Science Foundation, the U.S. Department of Energy, the National Aeronautics and Space Administration, the Japanese Monbukagakusho, the Max Planck Society, and the Higher Education Funding Council for England. The SDSS Web Site is http://www.sdss.org/.


\end{document}